\documentclass[journal,twocolumn,letterpaper]{IEEEJERM}

\ifCLASSINFOpdf

\else

\fi

\usepackage{times,amsmath,epsfig}
\usepackage{fancyhdr}
\usepackage{amsmath}
\usepackage{amsfonts}
\usepackage{amssymb}
\usepackage{array}
\usepackage{graphicx}
\usepackage{url}
\usepackage{subfigure}
\usepackage{bm}
\usepackage{breqn}
\usepackage{xcolor}
\usepackage{soul}
\usepackage{amssymb}
\usepackage[breaklinks=true]{hyperref}
\usepackage{breakcites}

\usepackage{booktabs}  % 提供更漂亮的表格线

\begin{document}

\vspace{1 cm}

\title{Multi-Center Study on Deep Learning-Assisted Detection and Classification of Fetal Central Nervous System Anomalies Using Ultrasound Imaging}

%医生详细信息未加

\author{Yang Qi$^1$, %~\IEEEmembership{Student~Member,~IEEE,}
        Jiaxin Cai$^2$,
        Jing Lu$^3$, 
        Runqing Xiong$^4$,
        Rongshang Chen$^1$, 
        Liping Zheng$^4$,
        Duo Ma$^4$
        %Majd Soudan$^2$, 
       
}% <-this % stops a space

\markboth{---}
{Z. Peng \MakeLowercase{\textit{et al.}}: Demo of IEEE Journal of Electromagnetics, RF and Microwaves in Medicine and Biology (JERM)}

\twocolumn[
\begin{@twocolumnfalse}
  
\maketitle

\begin{abstract}
 Prenatal ultrasound evaluates fetal growth and detects congenital abnormalities during pregnancy, but the examination of ultrasound images by radiologists requires expertise and sophisticated equipment, which would otherwise fail to improve the rate of identifying specific types of fetal central nervous system (CNS) abnormalities and result in unnecessary patient examinations. We construct a deep learning model to improve the overall accuracy of the diagnosis of fetal cranial anomalies to aid prenatal diagnosis. In  our collected multi-center dataset of fetal craniocerebral anomalies covering four typical anomalies of the fetal central nervous system (CNS): anencephaly, encephalocele (including meningocele), holoprosencephaly, and rachischisis, patient-level prediction accuracy reaches 94.5\%, with an AUROC value of 99.3\%. In the subgroup analyzes, our model is applicable to the entire gestational period, with good identification of fetal anomaly types for any gestational period. Heatmaps superimposed on the ultrasound images not only provide a visual interpretation for the algorithm but also provide an intuitive visual aid to the physician by highlighting key areas that need to be reviewed, helping the physician to quickly identify and validate key areas. Finally, the retrospective reader study demonstrates that by combining the automatic prediction of the DL system with the professional judgment of the radiologist, the diagnostic accuracy and efficiency can be effectively improved and the misdiagnosis rate can be reduced, which has an important clinical application prospect. All code is available at https://github.com/xiaqi7/Fetal-ultrasound.
\end{abstract}

% % Note that keywords are not normally used for peerreview papers.

\end{@twocolumnfalse}]

% % Put footnotes here
{
  \renewcommand{\thefootnote}{}%
  \footnotetext[1]{$^1$School of Computer and Information Engineering, Xiamen University of Technology}
   \footnotetext[2]{$^2$School of  Mathematics and Statistics, Xiamen University of Technology}
   \footnotetext[3]{$^3$Department of Obstetrics, Xiamen University Affiliated First Hospital}
   \footnotetext[4]{$^4$Department of Ultrasound Imaging, Xiamen Medical College Affiliated Second Hospital}
  %\footnotetext[2]{$^2$footnote (school)}
  %\footnotetext[3]{$^*$footnote (school)}
\footnotetext{This work is supported by the Natural Science Foundation of Fujian Province (2023J05083,2022J011396,2023J011434) and  Xiamen University of Technology Postgraduate Science and Technology Innovation Program Project (YKJCX2023061). Yang Qi, Jiaxin Cai, Jing Lu, Runqing Xiong and Rongshang Chen are co-first authors. Corresponding author: Jiaxin Cai (Email: caijiaxin@xmut.edu.cn).  }
}

\IEEEpeerreviewmaketitle

\section{Introduction}

Ultrasonography is popular as a non-invasive and radiation-free prenatal diagnostic method for its convenience and low cost \cite{chitty1991effectiveness}.
Antenatal ultrasound is a crucial imaging tool during pregnancy. It not only assesses fetal growth and development and detects congenital anomalies, but also provides important diagnostic information and support to clinicians through detailed imaging of the fetus and its associated structures \cite{sonek2007first}.
In ultrasound, physicians can assess the presence of congenital anomalies in the fetus with the help of two-dimensional (2D) and three-dimensional (3D) imaging, thus helping to significantly reduce the incidence of congenital disabilities. However, fetal ultrasound still faces some challenges in clinical practice, such as high fetal mobility, excessive abdominal wall thickness in pregnant women, and differences in interpretation between different observers \cite{xiao2023application}.
In fetal ultrasound, the acquisition of correct fetal position and standard planes relies on the expertise of the technician, and even experienced technicians find it time-consuming and laborious to take manual measurements such as head circumference (HC), biparietal diameter (BPD), and occipitofrontal diameter (OFD). Optimizing the prenatal ultrasound diagnosis process can significantly reduce the workload of the sonographer; therefore, the application of artificial intelligence (AI) and deep learning (DL) techniques in ultrasound imaging can significantly speed up the prenatal examination process while improving the accuracy and consistency of the diagnosis.

Artificial Intelligence  is widely used in medicine today \cite{rajkomar2019machine}. Deep learning, a subset of AI,  automatically extracts features from large amounts of data and performs efficient pattern recognition and prediction using deep neural network models \cite{suzuki2017overview}.
Deep learning has been widely applied in medical image processing, including the analysis of ultrasound images \cite{liu2019deep}.
Convolutional neural networks (CNN) are widely used as one of the most powerful methods in deep learning \cite{krizhevsky2012imagenet}. CNN is capable of extracting complex features from medical images for disease detection, diagnosis, and monitoring, greatly improving the accuracy and efficiency of medical image analysis \cite{yap2017automated}.
CNN is very promising in the detection of standard fetal ultrasound plane. Yu et al. \cite{yu2017deep} proposed a deep convolutional neural network (DCNN)-based method for automatic identification of fetal facial standard planes (FFSPs), which solves the problem of lack of performance of the traditional handcrafted feature methods in the identification of FFSP.
Chen et al.  \cite{chen2017ultrasound} used a composite architecture of Convolutional Neural Networks (CNN) and Recurrent Neural Networks (RNN) \cite{graves2013speech} for in-plane and cross-plane feature learning.
However, most studies using AI for fetal imaging have focused on the identification of normal fetal structures, and few have used AI to classify and diagnose images of fetal congenital malformations \cite{gong2019fetal,arnaout2021ensemble}.

Among the many fetal ultrasound planes of detection, the identification of fetal abnormalities is particularly important, as it is directly related to early diagnosis and timely treatment, which is of great significance for protecting fetal health and improving birth quality.
In particular, the development of head and brain structures plays a decisive role in the lifelong neurological function of the individual. Any impairment of head growth and neurodevelopment during the fetal period may lead to cerebrocranial malformations in the infant \cite{connors2008fetal}.
Therefore, by accurately recognizing abnormalities, doctors can detect potential health problems as early as possible, providing a valuable window of time for subsequent medical interventions. For example, Lin et al. \cite{lin2022use} proposed a real-time artificial intelligence-assisted image recognition system called PAICS (Prenatal Ultrasound Diagnosis Artificial Intelligence Conduct System). The system is designed to detect nine specific abnormality patterns in the standard ultrasound reference plane for intracranial malformations in the fetus. 
Xie et al. \cite{xie2020using} used CNN algorithm to segment the fetal cranial region and classified the fetal cranial images as normal or abnormal.
Chowdhury et al.\cite{chowdhury2023stackfbas}  proposed StackFBAs, a CNN-based framework that incorporates a stacking strategy, to detect abnormalities of the fetal brain from fetal images.

One of the most common congenital fetal abnormalities is the malformation of the central nervous system (CNS) of the fetal fetus \cite{paladini2007sonographic}.
These malformations, which can affect the brain, spinal cord, and their associated structures, are diverse and most have serious implications for fetal survival and development.
We use deep learning to conduct an in-depth study of four of these anomalies, namely anencephaly, encephalocele (including meningocele), holoprosencephaly, and rachischisis, which are some of the most common and typical types of fetal central nervous system (CNS) anomalies. These malformations are largely representative of the majority of developmental abnormalities of the fetal CNS. Anencephaly, encephalocele (including meningocele), and holoprosencephaly are fetal craniocranial anomalies that arise from defects in the closure of the neural tube or in the development of the brain during embryonic life.
% Anencephaly is a severe absence of the brain and skull, and the condition can be diagnosed by assessing whether the skull appears to be deficient above the level of the orbit and whether there is a protrusion of neural tissue from the cranial opening leading to the cranial defect.
% encephalocele (including meningocele) presents as an abnormal mass in the head with a distinct sac-like or pouch-like structure protruding outward from the skull defect, with cerebrospinal fluid or brain tissue visible inside. Common features include localized calvarial defects, herniation of brain tissue through the defect, and are often associated with ventriculomegaly and signs of increased intracranial pressure (such as dilatation of the lateral ventricles or displacement of the tentorium cerebelli).
% Holoprosencephaly is a severe congenital abnormality of central nervous system development, in which the embryonic cerebral hemispheres fail to split into two separate hemispheres. In prenatal ultrasonography, common characteristics include facial and cranial abnormalities (such as cyclopia, nasal aplasia, frontal concavity), cerebral structural anomalies (such as holoprosencephaly, agenesis of the corpus callosum, thalamic fusion), and lateral ventricle morphological anomalies \cite{paladini2018ultrasound}.
Anencephaly is characterized by the severe absence of the brain and skull, diagnosed by cranial defects above the orbital level and protruding neural tissue. Encephalocele (including meningocele) presents as a cystic or sac-like structure protruding from a cranial defect, containing cerebrospinal fluid or brain tissue. Holoprosencephaly is a congenital anomaly where the embryonic forebrain fails to divide into two hemispheres, prenatal ultrasound commonly reveals facial and cranial abnormalities, as well as cerebral structural anomalies \cite{paladini2018ultrasound}.
Analyzing the shape of the fetal skull is also one of the most important ways to diagnose craniosynostosis, such as the ‘Lemon Sign’, which resembles a lemon when viewed from the side due to the concave forehead and flattened posterior cranial fossa. Scaphocephaly is an anterior-posterior elongation of the head, which is boat-shaped or elongated, etc. On prenatal ultrasound, rachischisis presents as a widespread failure to close the entire spinal region, with the spinal cord and its overlying membranes markedly exposed, forming a distinctive sac or protruding structure. In addition, it is often accompanied by other neurological abnormalities, such as the lemon sign and banana cerebellum sign, as well as hydrocephalus \cite{nicolaides1986ultrasound}.
Rachischisis and other neural tube defects occur during early embryonic closure of the neural tube. The neural tube eventually develops into part of the central nervous system, including the brain, spinal cord, and their protective structures (such as the skull and spine). If the neural tube fails to close correctly in a particular area, it may result in different types of defects such as anencephaly, encephalocele, or rachischisis. These defects stem from the same developmental problems although they primarily affect different parts of the body. Rachischisis, exemplified by meningomyelocele, is frequently accompanied by Arnold-Chiari malformation type II \cite{kunpalin2021cranial}. This malformation involves caudal herniation of the cerebellar tonsils into the spinal canal, which may result in displacement of the brainstem and the fourth ventricle. This suggests that rachischisis may be accompanied by structural abnormalities of the brain that affect the normal anatomy of the cranium. Thus, they directly or indirectly affect the morphology and function of the cerebral cranium and are an important manifestation of abnormal CNS development.

We chose to investigate malformations of the fetal central nervous system (CNS) not only due to their high prevalence but also because early identification and intervention are critical to improving the prognosis and reducing long-term effects. By focusing on this area, our studies aim to provide new tools and methods for the early detection and diagnosis of fetal CNS malformations, using deep learning to bring substantial improvements to clinical practice. They are expected to help address the shortage of sonographers for basic prenatal ultrasound not only in China but also around the world, especially in less developed regions. In China, training a qualified sonographer is not only costly but also requires a long period of specialized training. Typically, it takes at least 3 to 5 years after graduation for a doctor to reach a level of proficiency, and even more time and effort is required to become an expert in the field. This long training process is not only time-consuming but also puts a huge strain on medical resources. In addition, the application of this technology can help reduce the financial burden on patients, especially in areas with limited resources, as automated systems can provide more efficient and cost-effective services.
Using deep learning algorithms, automated detection and classification of cranial anomalies in the fetal brain can be achieved, thus reducing the workload of sonographers and improving the consistency and reliability of diagnosis. As the algorithm continues to be optimized and the dataset expands, we expect the deep learning algorithm to be able to handle even more complex tasks, further improving its accuracy and utility. In the long run, through continuous technological improvement and clinical validation, the AI-based fetal ultrasound image analysis system is expected to become an important tool for prenatal screening, which not only improves the quality of diagnosis but also facilitates rational allocation of healthcare resources and ultimately improves the health of mothers and babies globally.

\vspace*{1 cm}

\section{Results}

\subsection*{Patient data}
\begin{figure}[!t]
\centerline{\includegraphics[width=1\columnwidth]{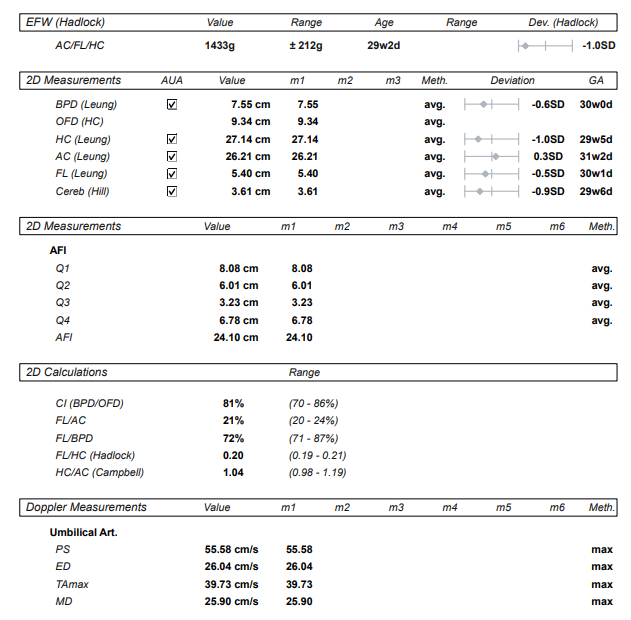}}
\caption{The ultrasound report indicates an estimated fetal weight (EFW) of 1433 grams at a gestational age of 29 weeks and 2 days. Key measurements include a biparietal diameter (BPD) of 7.55 cm, head circumference (HC) of 27.14 cm, abdominal circumference (AC) of 26.21 cm, femur length (FL) of 5.40 cm, and cerebellar diameter (Cereb) of 3.61 cm. The amniotic fluid index (AFI) is 24.10 cm, indicating normal amniotic fluid volume. Doppler measurements of the umbilical artery show a peak systolic velocity (PS) of 55.58 cm/s, end-diastolic velocity (ED) of 26.04 cm/s, time-averaged maximum velocity (TAmax) of 39.73 cm/s, and mean velocity (MD) of 25.90 cm/s.}
\label{bingli}
\end{figure}

We collected a dataset of abnormal pregnant women with fetal central nervous system anomalies from Xiamen University Affiliated First Hospital. Our dataset for model training and evaluation consisted of 1,662 fetal ultrasound images and 699 ultrasound videos. To make full use of the information in the videos, we use a computer script to convert these videos to still images frame by frame. To ensure that each frame clearly shows the abnormal state of the fetus and is distinct from the previous frame, we extracted one image every 80 frames from the start frame to the end frame. By this method, we extracted 5038 still images from the video. Eventually, as shown in Table \ref{data}, we obtained a dataset containing 6700 anomaly images (1662 original images plus 5038 images extracted from the video).
These fetal images were derived from the ultrasound findings of 37 pregnant women. The types of anomalies diagnosed in the fetus of each pregnant woman's womb included four categories: rachischisis, encephalocele (including meningocele), holoprosencephaly, and anencephaly. These four types of anomalies are representative of several types of fetal craniocerebral anomalies, all of which are neural tube defects or developmental anomalies with varying characteristics and degrees of severity. 
These pregnant women underwent ultrasound examinations at  Xiamen University First Affiliated Hospital   between 2019 and 2023. As shown in Table \ref{biao-binli}, in addition to ultrasound images, we collected data from 19 different pregnant women, as well as radiological and pathological reports related to the fetuses. For example, in a report for gestational age of 29 weeks and 2 days shown in Fig. \ref{bingli}, the ultrasound report provides a comprehensive set of fetal measurements, including the estimated fetal weight (EFW), biparietal diameter (BPD), head circumference (HC), amniotic fluid index (AFI), and Doppler measurement data, among other indicators.
To validate the classification performance of the model, we also collected prenatal ultrasound images of normal pregnant women from Xiamen Medical College Affiliated Second Hospital. This additional dataset contains four basic standard section images of mid-trimester fetuses, namely, thalamic transverse section, lateral ventricle transverse section, cerebellar transverse section, and spinal longitudinal section, with a specific number of 3207 images. We mixed the normal fetal dataset of Xiamen Medical College Affiliated Second Hospital with the abnormal dataset of Xiamen University Affiliated First Hospital for more comprehensive model validation.
The reference number of ethical approval is 2021052, adhering to the principles of the Declaration of Helsinki.
The feature extraction and pre-processing methods for all datasets are described in detail in the Materials and Methods section.

\begin{table*}
    \centering
    \begin{tabular}{>{\centering\arraybackslash}p{0.12\linewidth}>{\centering\arraybackslash}p{0.15\linewidth}>{\centering\arraybackslash}p{0.25\linewidth}>{\centering\arraybackslash}p{0.15\linewidth}>{\centering\arraybackslash}p{0.15\linewidth}}
    \toprule
         Dataset& Source of dataset&Content of dataset&  Practice&  Result\\
         \hline
 & & & &\\
         
         1662 fetal ultrasound images & Xiamen University Affiliated First Hospital & Ultrasound images of the cranium and other body parts of fetuses diagnosed with anencephaly, encephalocele (including meningocele), holoprosencephaly, rachischisis &  Crop out excess distracting information from the image&  1662 processed ultrasound images of various aspects of fetuses diagnosed with cerebral-cranial anomalies\\
 &  && &\\
         699 fetal ultrasound videos & Xiamen University Affiliated First Hospital & Ultrasound videos from fetuses diagnosed with anencephaly, encephalocele (including meningocele), holoprosencephaly, rachischisis &  Convert video to still images frame by frame &   5038 processed ultrasound images of various aspects of fetuses diagnosed with cerebral-cranial anomalies\\
 &  && &\\
          3207 ultrasound images of standard slices of fetus & Xiamen Medical College Affiliated Second Hospital  & Fetal cranial ultrasound images including thalamus in transverse section, lateral ventricle in transverse section, cerebellum in transverse section and spine in sagittal section&  Crop out excess distracting information from the image &  3207 processed ultrasound images of standard slices of normal fetuses\\
           \bottomrule
    \end{tabular}
    \caption{Processing and preparation of fetal ultrasound image and video datasets.}
    \label{data}
\end{table*}

\begin{table*}
    \centering
    \begin{tabular}{>{\centering\arraybackslash}p{0.16\linewidth}>{\centering\arraybackslash}p{0.15\linewidth}>{\centering\arraybackslash}p{0.08\linewidth}c}
    \toprule
    
         Types of neurocranial abnormalities&   Information for pregnant women&  GA at examination (weeks) & Deviation\\
         \hline
         &  2022041301 1988&  11w6d
 &CRL(-1.1SD)\\
         Anencephaly&  2023041305
 1987&  11w6d &CRL(-0.7SD)\\
         &  2023050507  1983&  13w1d
 &CRL(+1.1SD)\\
\hline
 & 2022051703 1996& 12w2d
 &-\\
 Encephalocele & 2022021807
 1988& 13w5d &BPD(-1.4SD)\\
 (including Meningocele)& 2021123002 1991& 15w3d
 &BPD(-2.9SD),HC(-3.5SD)\\
 & 2022051902 1990& 18w3d &BPD(+1.1SD),HC(+0.9SD),AC(+1.3SD),FL(+0.9SD)\\
 \hline
 & 2023040406 1989& 12w6d
 &-\\
 & 2022110901 1980& 13w1d &BPD(-0.9SD)\\
 & 2022090504 1987& 13w2d &-\\
 Holoprosencephaly& 2022120105
 1995& 13w6d
 &BPD(-1.6SD),CRL(+0.4SD)\\
 & 2022050703 1991& 17w1d
 &BPD(-1.4SD),HC(-2.1SD),AC(+0.5SD),Cereb(+1.4SD)\\
 & 2023041705
 1998& 22w1d &BPD(-1.3SD),HC(-2.5SD),AC(-1.5SD),FL(-1.5SD),Cereb(+0.6SD)\\
 & 2022092905 1988& 23w1d &BPD(-1.4SD),HC(-2.4SD),AC(+1.0SD),FL(+1.3SD),Cereb(+0.7SD)\\
 & 2022111603
 1987& 30w2d &BPD(-0.6SD),HC(-1.0SD),FL(-0.5SD),Cereb(-0.9SD)\\
 \hline
 & 2022070401 1979& 12w3d
 &BPD(+0.4SD),CRL(+0.7SD)\\
 Rachischisis& 2023032301
 1986& 18w3d &BPD(-0.7SD),AC(+0.4SD),FL(+1.3SD),Cereb(-1.6SD)\\
 & 2022103102 1995& 20w6d &BPD(-2.0SD),HC(-2.4SD),AC(-0.4SD),Cereb(-1.5SD)\\
 & 2023041402
 1998& 24w6d &HC(-0.5SD),FL(+0.4SD),Cereb(-0.7SD)\\
 \bottomrule
    \end{tabular}
    \caption{Ultrasonographic Measurements and Deviations from Standard Values in Fetuses with Various Neurocranial Anomalies. For each anomaly type, the table lists information on individual cases, specifying the gestational age (GA) at examination, and deviations from standard values for key biometric parameters such as crown-rump length (CRL), biparietal diameter (BPD), head circumference (HC), abdominal circumference (AC), femur length (FL), and cerebellar diameter (Cereb). These deviations are expressed in terms of standard deviations (SD) from the mean.}
    \label{biao-binli}
\end{table*}

\subsection*{Model Performance}

\begin{figure}[!t]
\centerline{\includegraphics[width=1\columnwidth]{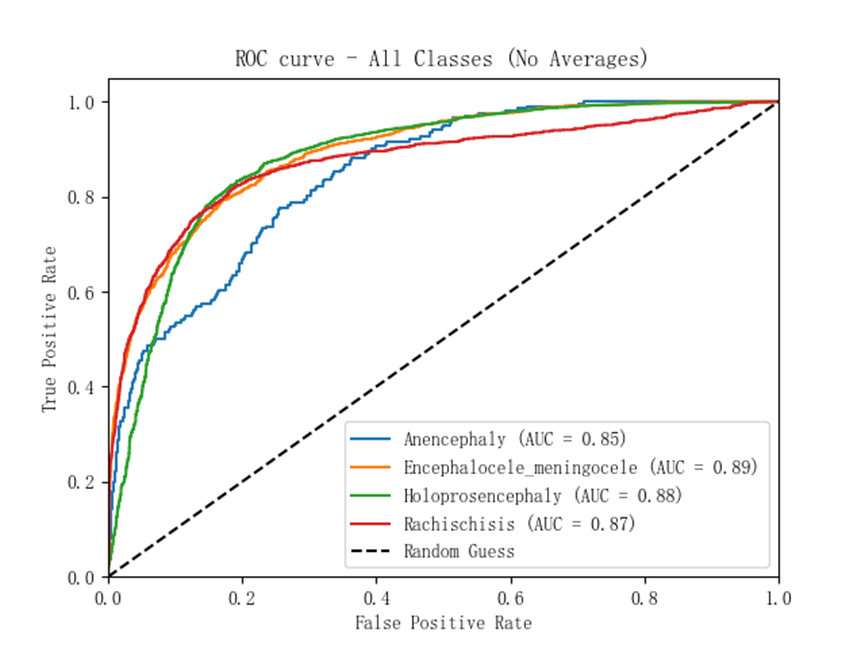}}
\caption{ROC curves for four categories of fetal anomalies.}
\label{4fenleiroc}
\end{figure}

Using a deep learning (DL) system to predict the fetal cerebral-cranial anomaly dataset, we first performed image-level testing. Specifically, we predicted the category of abnormality for each ultrasound image of the test pregnant women in each fold in the leave-one-out method of cross-validation. Eventually, we aggregated the prediction results for each image in all folds and obtained various performance metrics at the image level: accuracy of 73\%, precision of 60.43\%, recall of 65.7\%, F1 value of 61.8\%, and AUROC of 87.41\%.
In our experiments, we utilized subject operating characteristic (ROC) curves to assess the ability of the model to distinguish between anomalous cases. The ROC curves for four fetal anomaly categories, anencephaly, encephalocele (including meningocele), holoprosencephaly, and rachischisis, are presented in Fig. \ref{4fenleiroc}. The AUC values for each anomaly category were 0.85, 0.89, 0.88, and 0.87, respectively, indicating that the model has a high true-positive rate and low false-positive rate in detecting these anomalies. The use of AUC (area under the curve) as an assessment metric in the examination of central nervous system (CNS) abnormalities is due to its ability to comprehensively assess the overall performance of the model in distinguishing between abnormality-type scenarios at different thresholds, providing probabilistic predictions to support clinical decision-making.

\begin{figure}[!t]
\centerline{\includegraphics[width=1\columnwidth]{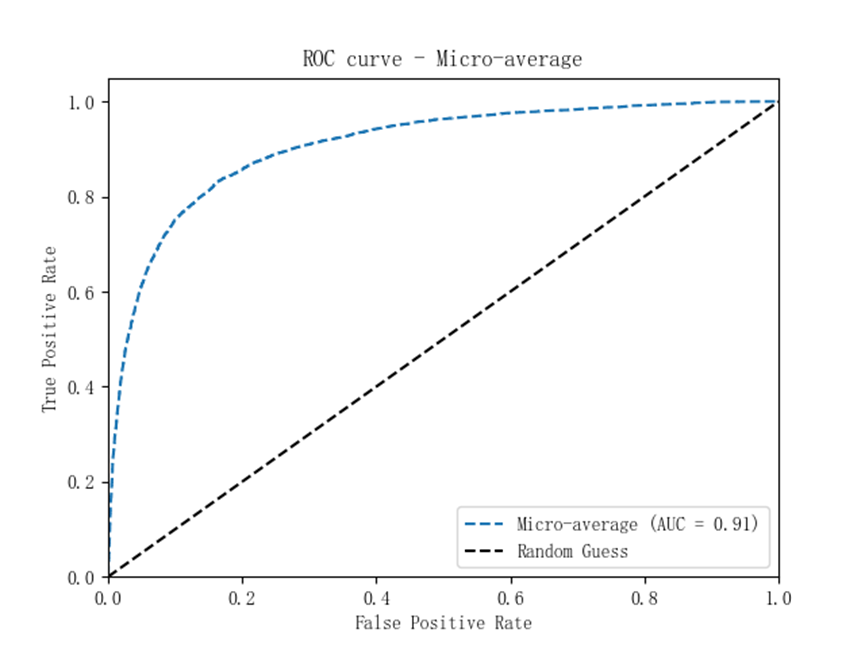}}
\caption{Micro-averaged ROC curves for four categories of fetal anomalies.}
\label{4fenleiwei}
\end{figure}

\begin{figure}[!t]
\centerline{\includegraphics[width=1\columnwidth]{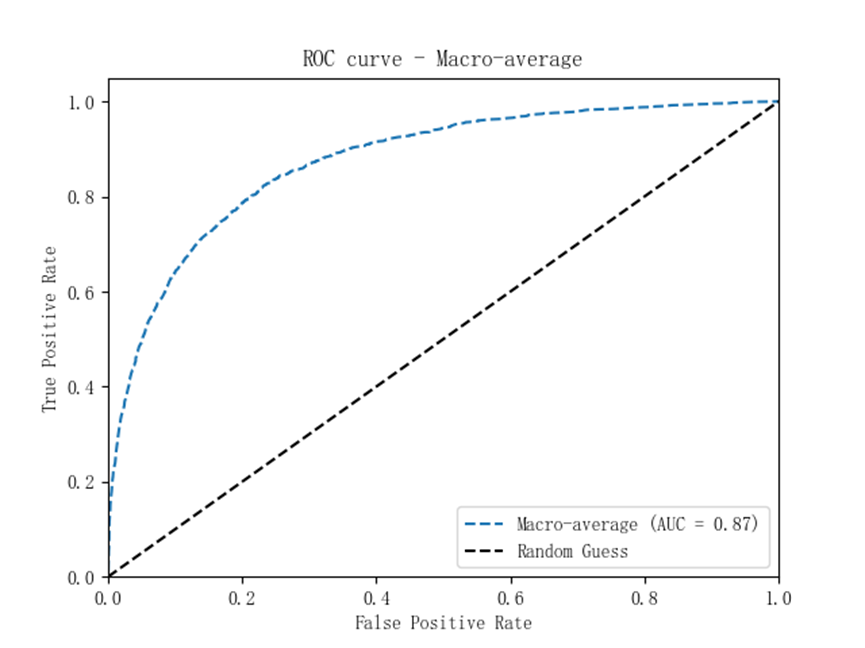}}
\caption{Macro-average ROC curves for four fetal anomaly categories.}
\label{4fenleihong}
\end{figure}

We also demonstrated the performance of the model in terms of CNS aberration checking using Micro-average ROC curves and Macro-average ROC curves. The Macro-average ROC curve focuses more on evaluating the average performance of the model on each category, while the Micro-average ROC curve focuses more on the overall performance. As shown in Fig. \ref{4fenleiwei} and Fig. \ref{4fenleihong}, the micro-average ROC curve (blue) has an AUC value of 0.91 and the macro-average ROC curve (blue) has an AUC value of 0.87, both of which are significantly higher than the random-guess value of 0.5, which suggests that the model has a high discriminatory ability overall. The micro-average ROC curve combines the true positive and false positive rates for all categories, whereas the macro-average ROC curve calculates the ROC curve separately for each category before averaging. Overall, the high AUC values reflect the ability of our model to maintain both a high true-positive rate and a low false-positive rate in identifying multiple fetal abnormalities. This means that the model is not only effective in detecting the actual presence of abnormalities but also reduces the likelihood of false positives, thus avoiding unnecessary further tests and patient anxiety.

\begin{figure}[!t]
\centerline{\includegraphics[width=1\columnwidth]{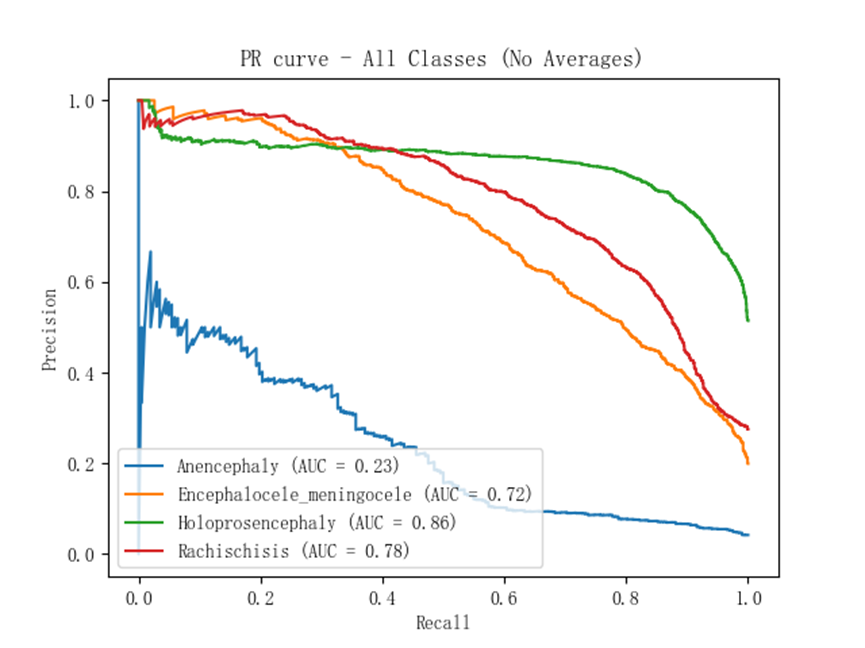}}
\caption{PR curves for four categories of fetal anomalies.}
\label{4fenleipr}
\end{figure}

In addition, as can be seen in Fig. \ref{4fenleipr}, the Precision-Recall (PR) curve for holoprosencephaly has the best performance, with an AUC value of 0.86, indicating that the model has high precision and recall in detecting this type of abnormality; rachischisis has an AUC value of 0.78, which also shows a good performance; in contrast, anencephaly has an AUC value of only 0.23, indicating that the model has a greater challenge in detecting this type of abnormality is more challenging. Because the number of anencephaly samples is extremely limited, it results in the model not being able to fully learn the features of this type of abnormality during training. These results have important implications for clinical practice: high precision reduces false positives and avoids unnecessary further tests and patient anxiety; high recall ensures that most actual anomalies are detected and reduces the risk of missed diagnosis. Therefore, by analyzing the PR curves, clinicians can better understand how the model performs in the detection of different types of fetal anomalies and set appropriate thresholds accordingly to optimize diagnostic decisions.

\begin{figure}[!t]
\centerline{\includegraphics[width=1\columnwidth]{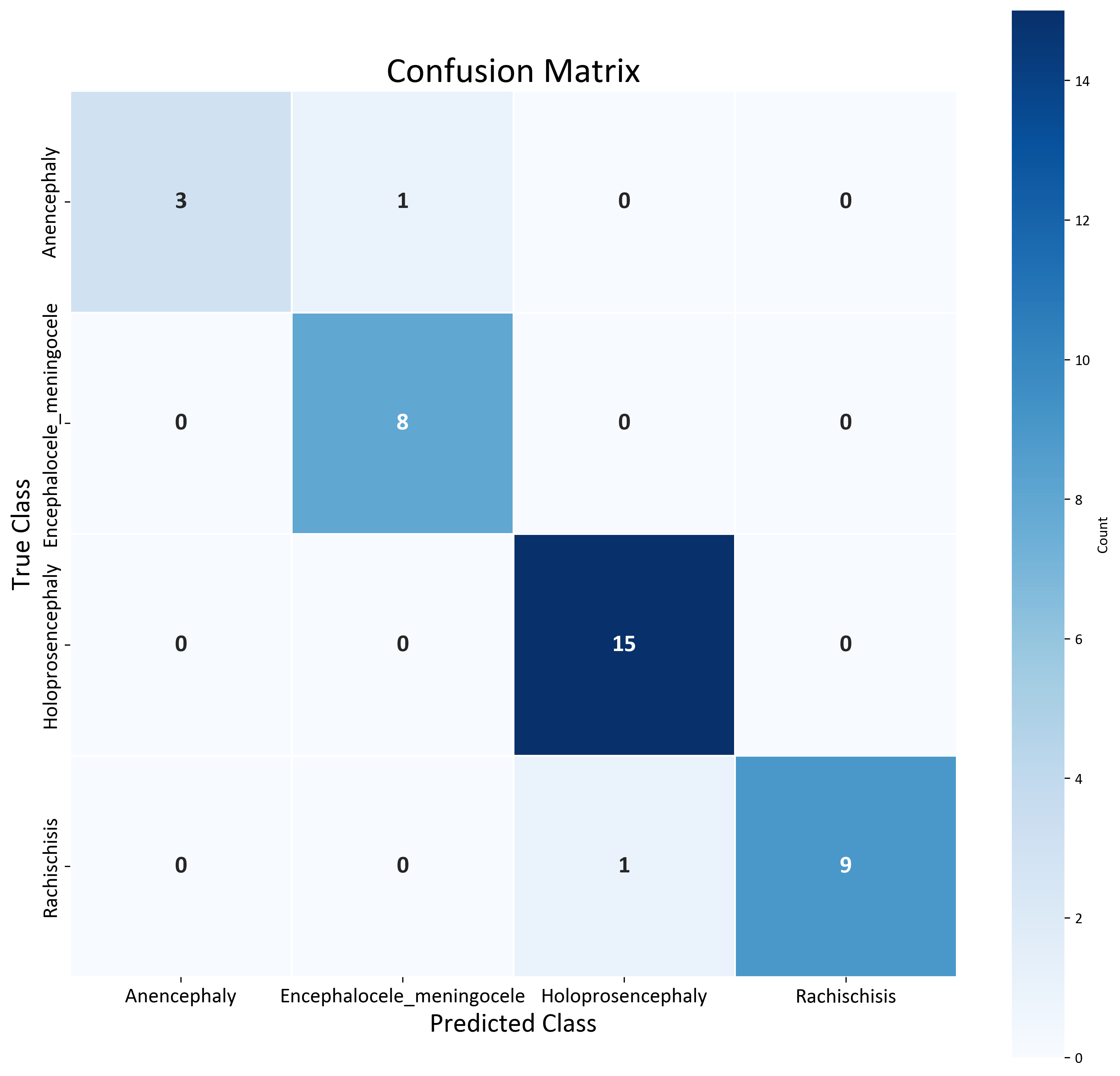}}
\caption{Individual-level confusion matrix for four fetal anomaly categories.}
\label{4flgthuan}
\end{figure}

However, image-level predictions are not the most expected results. In practical clinical applications, pregnant women undergoing ultrasound examinations usually do not make diagnostic conclusions based on a single ultrasound image of a suspected abnormality. Such a practice may lead to prenatal anxiety, psychological and physical discomfort, and may even lead to unnecessary treatment or termination of pregnancy. For this reason, a professional doctor performing a fetal ultrasound examination will usually combine multiple ultrasound images and videos for a comprehensive assessment to ensure the accuracy of the diagnosis.
To be closer to clinical practice, we further performed individual-level prediction. That is, the combined judgment of fetal abnormality categories was performed based on each complete ultrasound examination (including multiple ultrasound images and videos). In this case, the performance of our model is significantly improved with 94.5\% accuracy, 95.6\% precision, 91.2\% recall, and 92.8\% F1 value; our model performs well in the individual-level classification task. In addition, the AUROC is 99.3\%, and this near-perfect AUROC value implies that the model is able to distinguish normal and abnormal fetal cases almost perfectly. The specific confusion matrix is shown in Fig. \ref{4flgthuan}. The model successfully identified three cases of anencephaly, but one case was misdiagnosed as encephalocele/meningocele; all eight cases of encephalocele/meningocele were accurately identified; all 15 cases of holoprosencephaly were correctly classified; whereas, one of the nine cases of rachischisis was incorrectly classified as a holoprosencephaly. The model performed well in the detection of holoprosencephaly and rachischisis, with high true-positive and low false-positive rates, which is essential for reducing unnecessary further investigations, improving diagnostic efficiency, and providing more accurate prenatal counseling.
However, the misdiagnosis of anencephaly and rachischisis, especially the misidentification of anencephaly as encephalocele and rachischisis as holoprosencephaly, suggests that the algorithm needs to be further optimized to enhance its ability to differentiate between these similar conditions. Such improvements not only help to improve diagnostic accuracy, but also facilitate the effective implementation of early interventions, such as accurate diagnosis for anencephaly, which can better guide family decision-making, and timely surgical treatment for rachischisis, which can significantly improve the quality of life of the child. Therefore, continuous improvement of the model's performance is key to improving the overall quality of healthcare services and patient prognosis. In addition, enhanced multidisciplinary teamwork, combined with clinical experience and imaging features, can further improve the diagnostic accuracy of complex cases and ensure that each patient receives the best-individualised treatment plan.

Through individual-level prediction, we are able to more accurately identify specific categories of fetal abnormalities, thus providing more targeted guidance for prenatal and postnatal treatment. This approach not only improves diagnostic accuracy, but also reduces the false-positive rate, avoids unnecessary medical interventions and psychological burdens, and provides pregnant women and their families with more reliable and reassuring medical care.

In a further study, we extended the classification of fetal cerebral-cranial anomalies to five categories, including four types of anomalies: rachischisis, cerebral bulge (including meningeal bulge), holoprosencephaly, and anencephaly, and their corresponding normal cerebral-cranial sections (longitudinal section of the spinal column, lateral ventricle transverse section, thalamus transverse section, and cerebellum transverse section). This extension was designed to comprehensively assess the model's ability to differentiate between normal and abnormal samples, as well as its performance in accurately identifying specific types of abnormality in a wide range of CNS malformations. This approach significantly enhances the practical value and diagnostic accuracy of the model in real clinical applications, providing clinicians with a more reliable decision support tool.
Cross-sections of the lateral ventricles, thalamus, and cerebellum are important views for detecting and differentiating encephalocele (including meningocele), holoprosencephaly, and anencephaly. These views can provide critical anatomical information to help diagnose the three cerebral-cranial abnormalities mentioned above. For example, the thalamus cross-section reveals the structure of the thalamus and the third ventricle, which is useful in the diagnosis of holoprosencephaly; the lateral ventricle cross-section and cerebellum cross-section assess the shape and size of the lateral ventricles and the development of the cerebellum, reveal intraventricular changes accompanying encephalocele, and reveal abnormal cerebellar position or dysplasia, which are useful in identifying encephalocele (including pachymeningeal) and anencephalic babies.
Longitudinal views of the spine are used to detect neural tube defects such as rachischisis. This view clearly demonstrates the midline structure of the spine and its continuity, helping to identify open and closed rachischisis, herniation of spinal canal contents, and concomitant soft tissue abnormalities. In addition, the longitudinal view of the spine allows assessment of the location and morphology of the spinal cord and nerve roots, providing important information for diagnosing and differentiating between different types of rachischisis. Although rachischisis primarily affects the spine, it may also be accompanied by other cerebral-cranial abnormalities such as hydrocephalus or other neural tube defects. Therefore, in our study, longitudinal spinal sections were not only used to detect rachischisis, but were also included in the overall assessment system to comprehensively analyze possible combined cerebro-cranial abnormalities.

\begin{figure}[!t]
\centerline{\includegraphics[width=1\columnwidth]{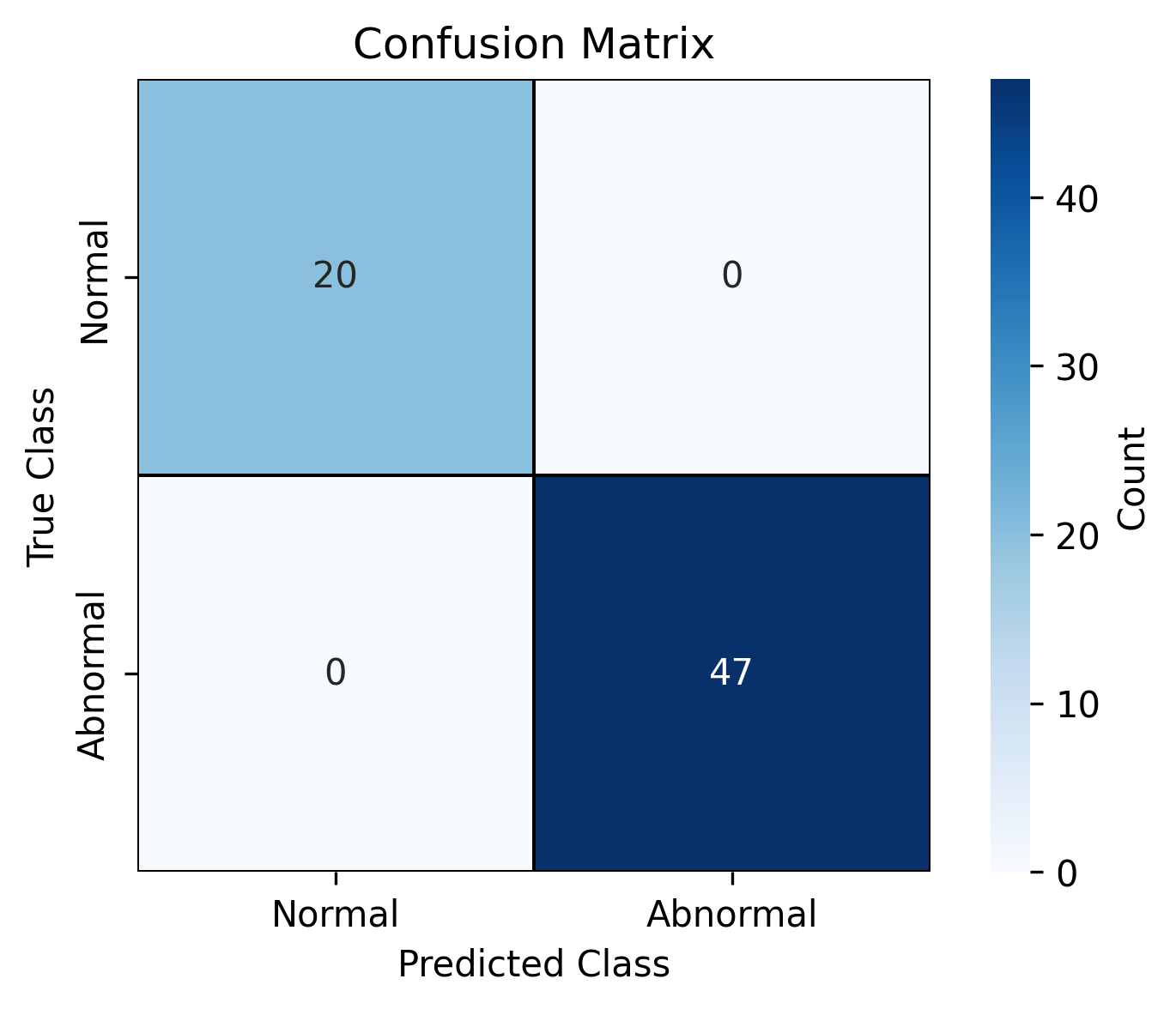}}
\caption{Individual-level confusion matrix for 2 fetal anomaly categories.}
\label{2flgthuan}
\end{figure}

As shown in Fig. \ref{2flgthuan}, this study evaluated the performance of the developed model in distinguishing between normal and abnormal fetal brain cranium samples. The experimental results show that the model demonstrates excellent ability in identifying whether there are structural abnormalities in the fetal brain cranium. The model not only efficiently distinguishes between normal and abnormal samples, but also achieves accurate classification in all test cases, demonstrating excellent classification performance. Importantly, the model was able to accurately identify all types of normal and abnormal samples without missing detection, which is potentially valuable in clinical applications.

\begin{figure}[!t]
\centerline{\includegraphics[width=1\columnwidth]{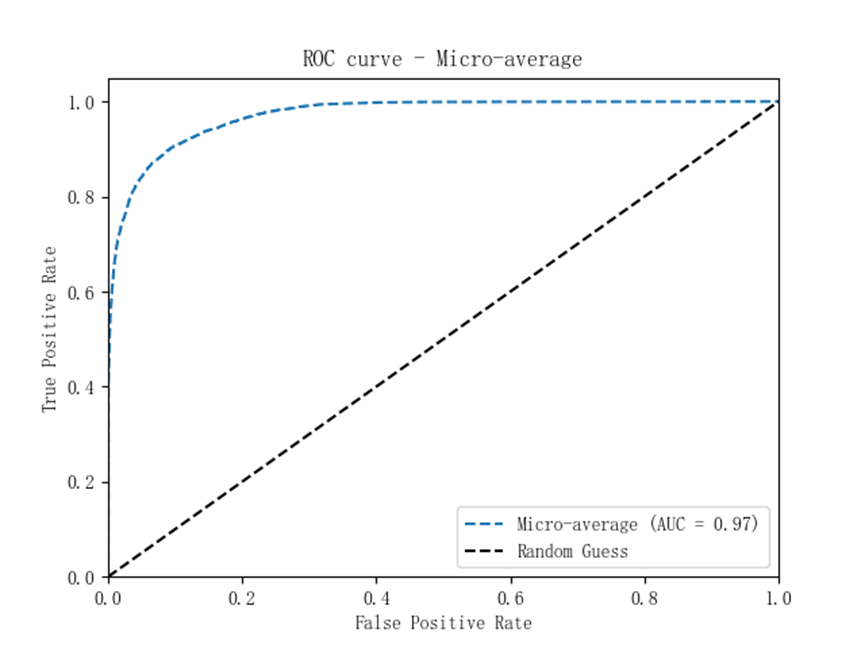}}
\caption{Picture-level micro-average ROC curves for 5 fetal anomaly categories.}
\label{5gtwei}
\end{figure}

\begin{figure}[!t]
\centerline{\includegraphics[width=1\columnwidth]{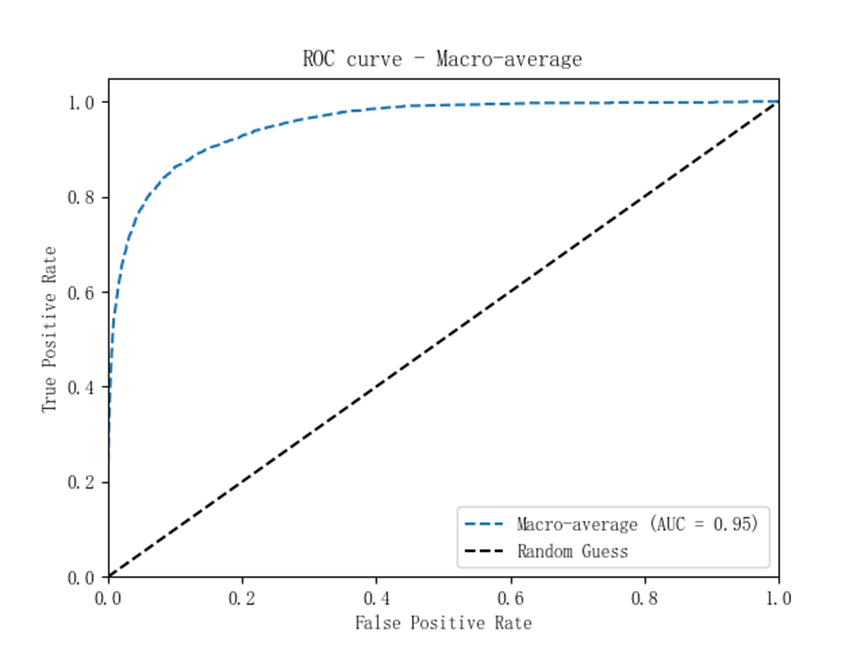}}
\caption{Picture-level macro-average ROC curves for 5 fetal anomaly categories.}
\label{5gthong}
\end{figure}

To further evaluate the classification performance of the model in a wide range of central nervous system (CNS) malformations and normal samples, we performed detailed analyzes with an accuracy of 98.25\%, a precision of 97.78\%, a recall of 98.67\%, and an F1 value of 98.13\%.
We used picture-level receiver operating characteristic (ROC) curves to evaluate the classification performance of the classifier. The area under the micro-averaged (micro-average) ROC curve (AUC) of the model is 0.97 and the area under the macro-averaged ROC curve (AUC) is 0.95, as shown in Fig. \ref{5gtwei} and \ref{5gthong}. This high AUC value suggests that the model has an excellent ability to distinguish between positive and negative samples.
Notably, all samples in the normal category were correctly classified; for the specific abnormality types, only 1 abnormal sample was misclassified as another abnormality type, but none of the abnormal samples were incorrectly classified as normal. This result is particularly important because the impact of misidentifying abnormal types as other abnormal types is relatively small in clinical practice. Although this situation may increase the workload of physicians by causing them to perform additional ultrasound examinations on pregnant women to confirm the specific type of abnormality, it ensures diagnostic accuracy and avoids the potential risks associated with misdiagnosis. In contrast, the risk of misidentifying an abnormal type as normal is significantly greater, which may lead to delays in diagnosis and treatment, which in turn may affect the health and development of the fetus. Therefore, the high sensitivity and specificity demonstrated by the model in this study not only effectively reduces the occurrence of underdiagnosis, but also provides strong support for clinical diagnosis and ensures the reliability and accuracy of diagnosis.

In the process of evaluating the model's classification performance for a wide range of central nervous system (CNS) malformations and normal samples, we found that the overall classification performance of the model was significantly improved after introducing an explicit normal class (e.g., normal brain cranial section) into the original four-classification task (rachischisis, encephalocele, holoprosencephaly, anencephaly) to form a five-classification task for a specific medical image classification task. This is because adding the normal category as a negative example, not only provides a clear control for the model to help it better learn the difference between abnormality and normal, but also increases the diversity of the data and the richness of the feature space, which in turn reduces inter-category confusion, alleviates the problem of category imbalance, and enhances the model's generalization ability. These findings suggest that deep learning techniques have significant potential for application in the field of prenatal diagnosis, especially in improving the accuracy of fetal cerebral-cranial anomaly detection. Future research can further explore how to optimize the model to better serve clinical practice and enhance the quality of patient care.

\begin{figure}[!t]
\centerline{\includegraphics[width=1\columnwidth]{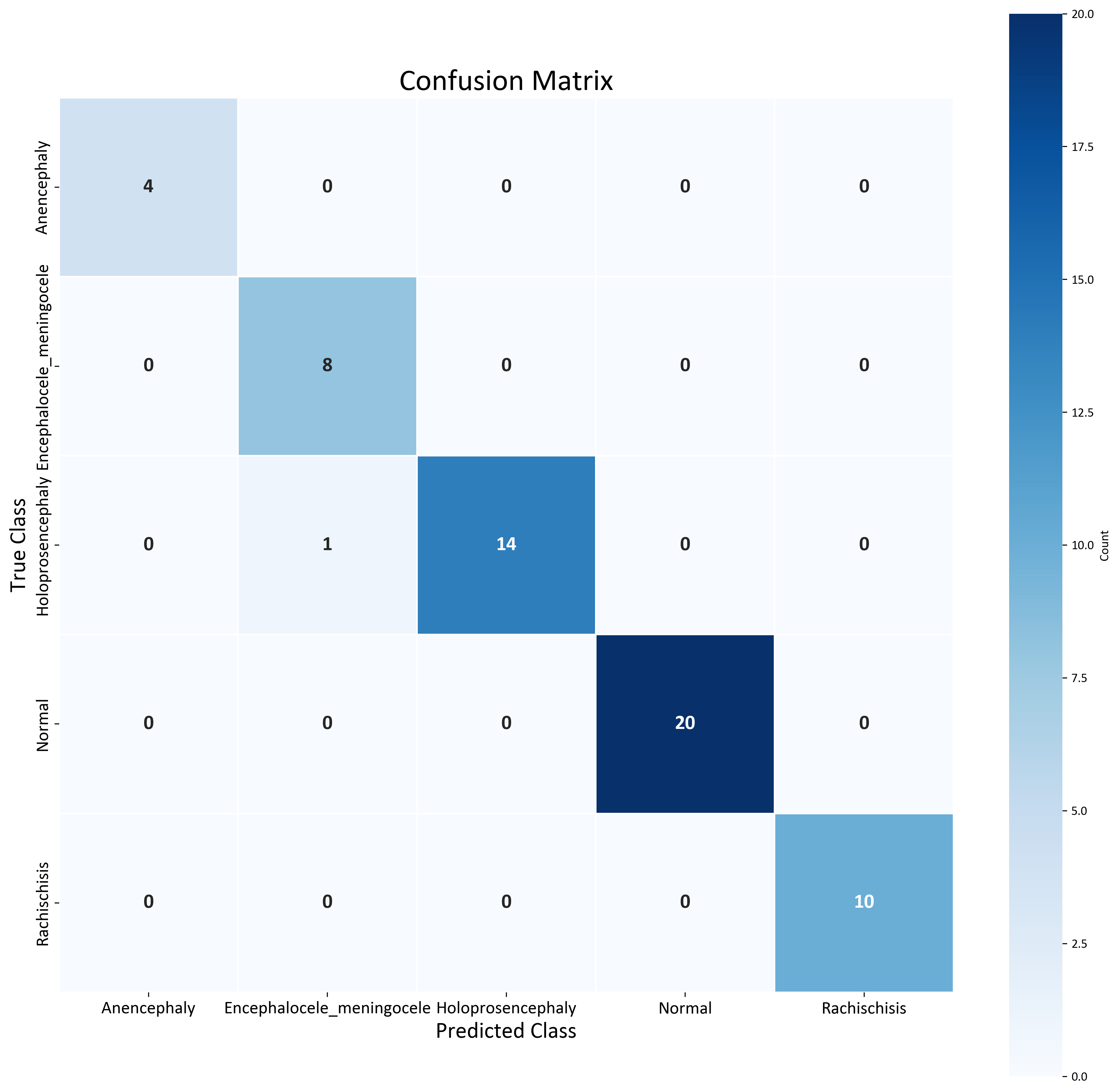}}
\caption{Individual-level confusion matrix for 5 fetal anomaly categories.}
\label{5gthuan}
\end{figure}

\subsection*{Subgroup analysis}

Subgroup analyzes were designed to determine how well the model performs in different assessment categories and to explore whether specific subgroups of pregnant women are more likely to benefit from the model. This assessment is clinically important for identifying subgroups of high-risk pregnant women and can help in the early detection and intervention of fetal anomalies, thereby improving prognosis. As shown in Table \ref{biao-binli}, we performed a detailed subgroup analysis of 19 labor reports, of which 5 were in early pregnancy ($\leq$12 weeks), 13 in mid-term pregnancy (13-28 weeks) and 1 in late pregnancy ($\textgreater$28 weeks). At early gestational stages, the fetus is small in size, many organs and structures are not yet fully developed, and brain structures are not sufficiently visible to be assessed in detail by ultrasound. For example, severe malformations such as anencephaly and encephalocele may not be easily recognized at early gestational stages. Early pregnancy reports focus on basic fetal growth indicators such as CRL and NT, which provide preliminary information but are not sufficient for a comprehensive assessment of all fetal structures. The vast majority of measurements in the five reports of early pregnancy were within the normal range, with only a very small number falling a little below the normal range, e.g., CRL measurements were below the normal range in two reports, which may suggest slower fetal growth or other potential problems. However, these indicators cannot determine whether there are abnormalities in the subsequent development of the fetus, and further anatomical ultrasound is required to rule out or detect other structural abnormalities, such as heart defects, neural tube defects, etc.

\begin{figure}[!t]
\centerline{\includegraphics[width=1\columnwidth]{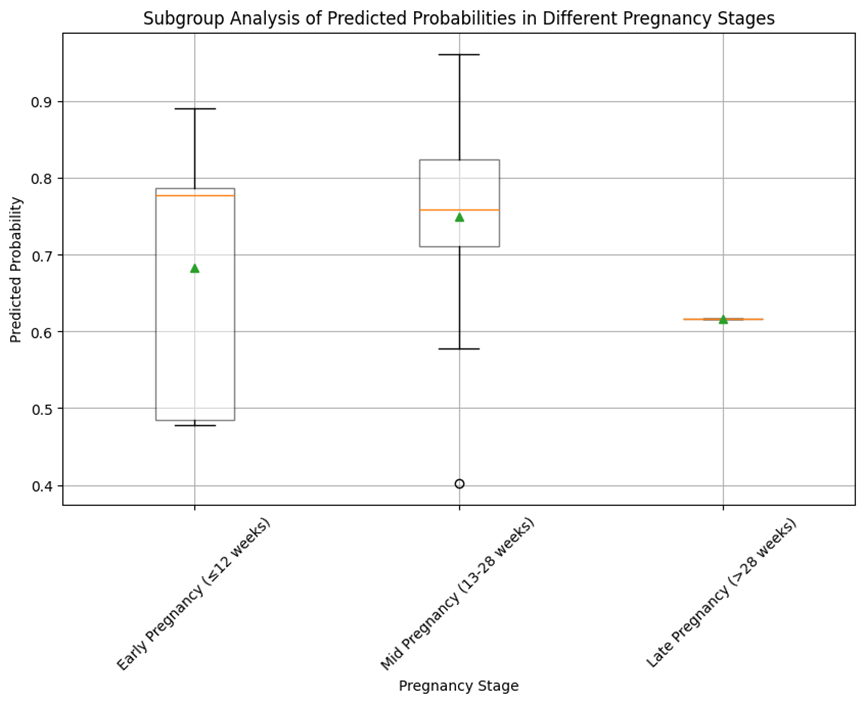}}
\caption{Predicted probability distribution of types of fetal anomalies in different trimesters.}
\label{yazu}
\end{figure}

As shown in Fig. \ref{yazu}, after model prediction, the type of abnormality was successfully predicted in all of these five early pregnancy reports, indicating that the model has strong abnormality detection capability. Compared to traditional reports, our model can accurately predict the presence of certain central nervous system (CNS) abnormalities in early pregnancy from ultrasound images, thus detecting the type of fetal abnormality at the early pregnancy stage, targeting treatment, and providing more opportunities and time to improve prognosis. Nonetheless, due to the difficulty in identifying the type of fetal abnormality in early pregnancy, even though the model determined that the highest probability score for the fetal abnormality category was 0.8906, two of the highest probability scores for the abnormality category were approximately only 0.48. This suggests that the model is still uncertain in identifying certain specific types of fetal abnormality in the early pregnancy stage. In contrast, at the mid-pregnancy stage, the highest probability anomaly type scores derived by the model were significantly higher, with an average score of 0.75. Ultrasound examinations at the mid-pregnancy stage can provide a great deal of information about fetal growth and development, and by this time, fetal organs and structures are more developed, allowing for a more detailed assessment of fetal health. In these 16 reports, there were more instances of abnormal parameters, and in some pregnancies, parameters such as BPD or HC were below the normal range in the maternity reports, but certain measurements in the abnormal range do not necessarily mean that the fetus has some kind of central nervous system (CNS) abnormality.
In only one of these 16 pregnancies did the predicted type of fetal abnormality not match the actual type of fetal abnormality. It is worth noting that in this sole case of misclassification, the probability score of misclassification and the probability score of correct classification of the model outputs were both 0.4028. This situation is not an obvious error, but rather a reflection of the challenges faced by the model in dealing with certain complex or ambiguous situations. In late pregnancy reports, detailed ultrasound examination and Doppler flow assessment revealed abnormalities in the vast majority of parameters, suggesting the presence of severe growth restriction in the fetus. The abnormal ranges found at this stage usually give a clearer indication of the type of fetal abnormality and therefore the model successfully predicts the correct class of abnormality.

\begin{figure}[!t]
\centerline{\includegraphics[width=1\columnwidth]{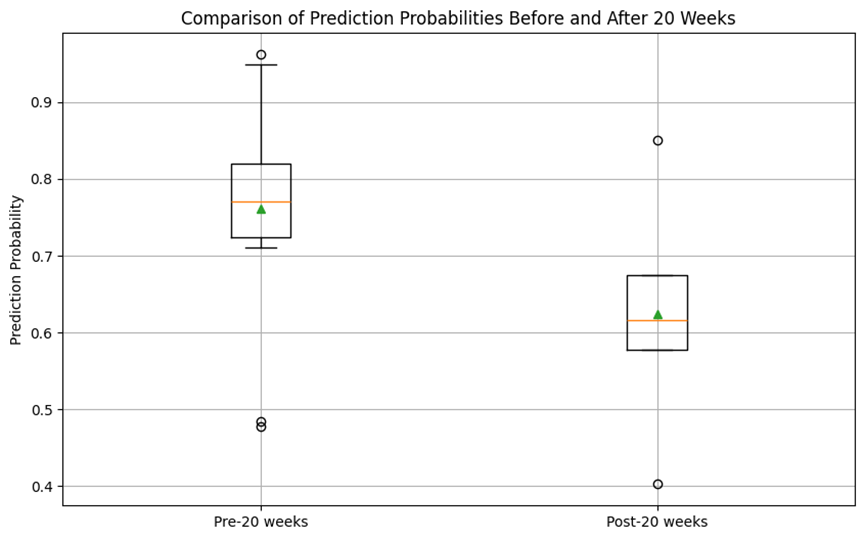}}
\caption{Predicted probability distribution of the types of fetal anomalies at different trimesters, using 20 weeks' gestation as a cut-off point.}
\label{tjx}
\end{figure}

Given that the optimal time for detailed anatomical screening (also known as ‘macrosomia’ or ‘structural screening’) is usually between 20 and 24 weeks of gestation, i.e., mid-pregnancy, we chose to statistically analyze the prenatal ultrasound data at 20 weeks of gestation, using 20 weeks of gestation as the cut-off point. Of the 19 deliveries, 14 were completed before 20 weeks, whereas 5 were performed after 20 weeks. As shown in Fig. \ref{tjx}, by comparing the prediction performance between the two groups, we obtained a p-value of 0.1432, which indicated that there was no statistically significant difference in the probability scores for predicting the true type of abnormality between the two groups ($p>0.05$).
This result suggests that our model performs consistently before and after the optimal time for detailed anatomical screening and is applicable throughout pregnancy. In other words, the model did not show statistically significant changes in its ability to predict the type of fetal abnormality, either within or outside the recommended screening window. This finding has important clinical implications, as it implies that the model can provide reliable diagnostic support over a wider timeframe and is not limited to a specific screening timeframe.

\subsection*{Lesion localization using heatmaps}

\begin{figure*}[!t]
\centerline{\includegraphics[width=1.4\columnwidth]{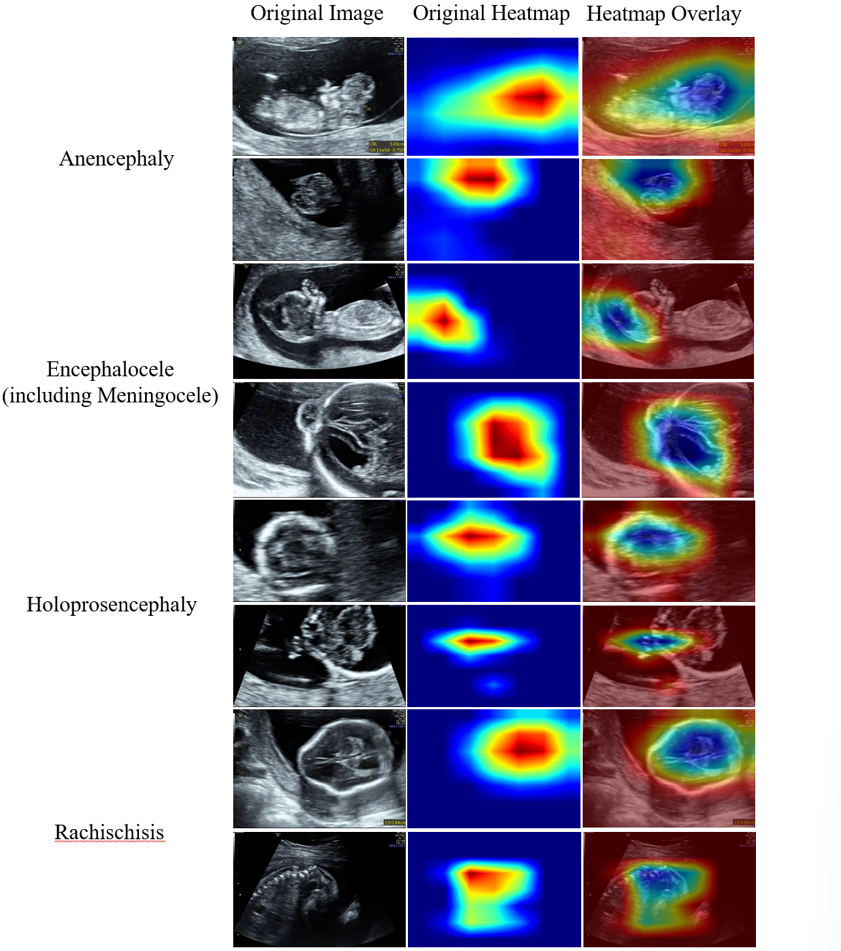}}
\caption{Stacked diagrams of different anomaly types. The eight original ultrasound images tested include four abnormal cases, two of each. The fetal ultrasound image on the left is the original test image; the image in the middle is the algorithm-generated heatmap; the image on the right is the heatmap superimposed on the original ultrasound image.}
\label{tu-reli}
\end{figure*}

Deep learning (DL) algorithms are often considered as ‘black box’ models and it is difficult to explain their internal decision-making process in detail. However, in our study, we achieved significant explainable results using heatmap and superimposed image techniques to assist in ultrasound image prediction of fetal cranial anomalies. We generated heatmaps using Grad-CAM (Gradient Weighted Class Activation Mapping), which determines which regions contribute the most to the model's prediction results by calculating the gradient of a particular class on the last layer of the feature map of a convolutional neural network (CNN). The generated heatmap is normalized and color mapped and overlaid with the original ultrasound image to form an intuitive visualization of the results. This is shown in Fig. \ref{tu-reli}, where the red color indicates the key regions that the algorithm mostly focuses on. In fetal brain cranial ultrasound images, heatmaps are effective in highlighting abnormal regions such as rachischisis, encephalocele (including meningocele), holoprosencephaly, and anencephalic lesions. By overlaying the heatmap with the original ultrasound image, physicians can more accurately localize and assess these abnormalities to determine whether the localization of key structures on the heatmap corresponds exactly to the actual area of the lesion. 
We selected two images from each category of abnormality in the test set to overlay with the generated heatmap and saw that in the detection of rachischisis, the heatmap clearly shows areas of the spine that are not closed; in the detection of encephalocele (including meningocele), the heatmap highlights the parts of the brain that are protruding from the brain tissue or the cerebrospinal fluid;  in the detection of the holoprosencephaly, the heatmap can help to identify the abnormal connections between the cerebral hemispheres; in the detection of anencephaly, the heatmap can clearly show areas where the cerebral cortex is missing. 
The experimental results show that the heatmap not only provides a visual explanation for the algorithm, but also provides important informational support to the physician by highlighting key areas that need to be reviewed. This phenomenon suggests that the primary role of heatmaps is to provide an intuitive visual aid to help physicians quickly identify and validate critical regions, and also to enhance their understanding of and trust in the algorithm's output. The use of heatmaps and superimposed images significantly improves the accuracy and interpretability of lesion detection, providing strong support for clinical diagnosis. We believe that heatmaps will play an increasingly important role in future differential diagnosis as technology continues to advance and clinical applications deepen.

\subsection*{Retrospective reader study}
In order to fully evaluate the prediction performance of our deep learning (DL) system and perform a detailed error analysis, we invited two experienced radiologists to participate in the study. These two doctors, radiologist A and radiologist B, who have 18 and 7 years of professional experience respectively, independently evaluated ultrasound images extracted from the fetal cranial anomaly dataset and the normal dataset. We aimed to systematically analyse the accuracy and reliability of the model and to explore its potential application in clinical practice.
Abnormal ultrasound images in the test set were randomly selected from the obstetric sonograms of the 37 mothers of abnormal fetuses to ensure that every pregnant woman had an image included in the test. Normal images were then randomly selected from the dataset of normal fetuses in equal numbers to the abnormal images.

\begin{figure}[!t]
\centerline{\includegraphics[width=1\columnwidth]{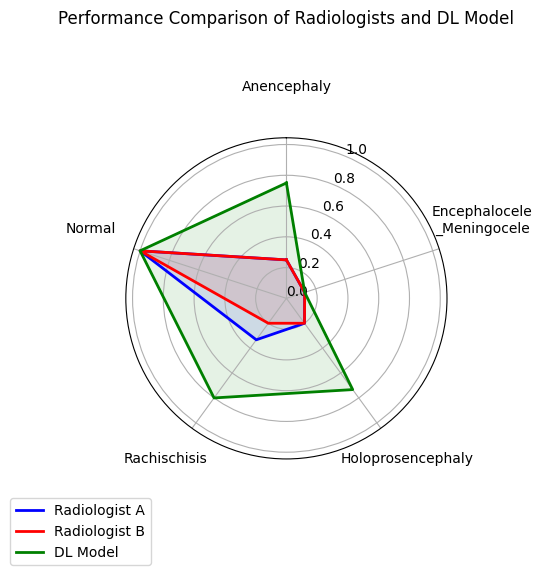}}
\caption{The predictions of the DL system were compared with the professional judgement of the radiologists based on recognition rate. Radiologists A and B are those with 18 and 7 years of professional experience, respectively.}
\label{leida}
\end{figure}

As shown in Fig. \ref{leida}, by comparing the prediction results of the DL system with the professional judgement of the two radiologists, we found that the DL system significantly outperformed the radiologists in the recognition rates of all four types of anomalies and normal types. In particular, the recognition rate on the three types of anomalies, namely anencephaly, holoprosencephaly and rachischisis, was close to 80\%, whereas the physicians' recognition rate was only about 25\%. For normal craniocranial views, the two reached almost exact agreement, with all test images correctly predicted, indicating that the DL system has an extremely high recognition rate for normal fetal craniocranial ultrasound images.
However, both the DL system and radiologists performed low on the identification of the type of abnormality, encephalocele (including meningocele). This may be due to the lack of distinctive fetal anomalies in the randomly selected ultrasound images from the fetal maternity pictures and the limited amount of test data. In practice, diagnosis of fetal abnormalities requires more maternity pictures to improve the accuracy. As shown in Fig. \ref{5gthuan}, the DL model trained with a large amount of data also exhibits a high prediction rate for such abnormalities.

In order to improve the diagnostic accuracy in clinical applications, we propose to combine the DL system with the professional judgement of radiologists to form a hybrid prediction model. This combination not only exploits the high accuracy of the DL system, but also compensates for its limitations in complex cases. For example, for complex abnormal pictures with atypical morphological features or hidden lesions, it is sometimes difficult for the two doctors to accurately determine whether the fetus is abnormal or to accurately determine the specific type of abnormality, whereas at this time the prediction scores of the DL system can give the radiologists detailed prediction data in order to improve its prediction accuracy and reliability in complex cases.

Specifically, in this independent test between the DL system and radiologists, the DL system's predictions performed particularly well for a number of cases in which there was significant disagreement between the two physicians.
For example, our model predicted anencephaly with a high prediction score of 0.9316 in cases where the anencephaly was incorrectly identified as a normal fetus by radiologist A. When rachischisis was incorrectly identified as a normal fetus by radiologist B, our model predicted rachischisis with a prediction score of 0.2694, anencephaly with a prediction score of 0.2968, and encephalocele with a prediction score of 0.273. Although the prediction score for rachischisis was not the highest, it was similar to the prediction scores for the other anomaly types, suggesting that the model was able to identify the presence of an anomaly in the fetal cranium, but the exact type needs to be further confirmed. These results suggest that the prediction scores provided by the DL system can be used as an adjunct to help physicians perform a secondary assessment in cases of uncertainty when two radiologists misdiagnose certain anomalous images as normal.
By combining the automatic prediction of the DL system with the professional judgment of the radiologist, the accuracy and efficiency of diagnosis can be effectively improved and the misdiagnosis rate can be reduced. This hybrid prediction mode not only improves the diagnostic accuracy, but also enhances the confidence of clinicians, which has an important clinical application prospect.

% \begin{table*}
%     \centering
%     \begin{tabular}{>{\centering\arraybackslash}p{0.3\linewidth}>{\centering\arraybackslash}p{0.2\linewidth}>{\centering\arraybackslash}p{0.2\linewidth}}
%     \toprule
%          &  Radiologist&  DL system\\
%          \hline
%  & & \\
         
%          Identification  of normal fetal ultrasound images&  Radiologist A: All correct Radiologist B: All correct&  All correct\\
%  & & \\
%          \hline
%          &  &  \\
%  & Radiologist A: normal Radiologist B: Anencephaly& Anencephaly: 0.9316     Encephalocele: 0.0337     Rachischisis: 0.0265\\
%  Fetal cerebral-cranial anomalies in ultrasound images to identify divergent conditions& & \\
%          &  Radiologist A: Rachischisis Radiologist B: normal&      Anencephaly: 0.2968    Encephalocele: 0.2730     Rachischisis: 0.2694\\
% \bottomrule
%  & & \\
%     \end{tabular}
%     \caption{The predictions of the DL system were compared with the professional judgement of the radiologists. Radiologists A and B are those with 18 and 7 years of professional experience, respectively. The DL system predicts results by taking the three with the highest probability scores.}
%     \label{read}
% \end{table*}

% \begin{figure}[h]
% \centering
% \includegraphics[width=\linewidth]{./figures/beforeAfter.png}
% \caption{Caption }
% \label{fig:beforeAfter}
% \end{figure}

\section{Discussion} 

In this study, we constructed and validated a deep learning (DL) model-based prediction system for fetal cerebral-cranial anomalies, focusing on the automatic detection of central nervous system (CNS) abnormalities in ultrasound images. By training a dataset covering four typical fetal cerebral-cranial anomalies, including anencephaly, encephalocele (including meningocele), holoprosencephaly, and rachischisis, we constructed an efficient and accurate diagnostic model. 

Currently, there are relatively few studies using artificial intelligence (AI) to predict fetal cerebral-cranial anomalies, and most of the existing models are limited to binary classification (normal vs. abnormal) or identifying general intracranial image patterns \cite{lin2022use,xie2020using,chowdhury2023stackfbas}. In contrast, our approach not only distinguishes between normal and abnormal fetal cranial images but also accurately predicts specific types of abnormalities. It provides detailed and specific diagnosis for physicians, reduces their workload, especially in resource-constrained areas for prenatal screening, and has very high prediction accuracy.
The results of the study showed that the model performed in terms of classification accuracy (94.5\%) and AUROC value (99.3\%), implying that the model is able to efficiently differentiate between normal and abnormal cases and has a well-balanced predictive ability for each type of abnormality, which provides more reliable decision support for the clinic. 

Specific predictive outcomes enable families to receive targeted counselling to help them better understand potential outcomes and prepare for necessary postnatal treatment or interventions. It also facilitates early detection of abnormalities, which is critical for timely surgery or other treatment and may significantly improve the long-term prognosis of affected infants. The high precision and recall of our model minimizes unnecessary follow-up tests, reduces patient anxiety, and ensures that only truly abnormal cases receive further attention.
Our DL system outperforms humans in the prediction of fetal cerebral-cranial anomalies, especially in the prediction of cerebral-cranial anomalies with accuracy and speed far exceeding that of physicians.

Through subgroup analyzes, we found that the predictive ability of the DL system varied at different gestational stages. Especially in the middle gestation stage (13-28 weeks), the predictive performance of the model was significantly improved and the accuracy was significantly better than that in the early gestation stage ($\leq$12 weeks). In the early stage, the identification accuracy of the model is limited by the poor clarity of the ultrasound images because the fetal organs and structures are not yet fully developed. However, as the pregnancy progresses and the fetal structure becomes clearer, the model is able to extract key features and classify them more accurately.
Because of this, deep learning models still suffer from the ‘black box’ problem, i.e., their decision-making process lacks interpretability, which limits their clinical applications. To improve the transparency of the model, we use heatmaps to overlay the original ultrasound images to help clinicians visualize the basis of the model decisions. Heatmaps not only enhance the interpretability of the model but also improve the accuracy of lesion detection, which plays a key role, especially in complex cases and images with blurred edges. In addition, to address the problem of category imbalance (e.g., fewer samples of anencephalic children), we used a weighted cross-entropy loss function to improve the model's performance on a few categories.
Although the DL system in this study achieved significant results on several metrics, the model still needs to be combined with the judgment of experienced radiologists in practical applications. The study demonstrated that a DL system combined with physician judgment can significantly improve diagnostic accuracy by using the DL system as an auxiliary tool to help quickly locate and screen for abnormalities, thereby improving diagnostic efficiency and reducing misdiagnosis and underdiagnosis.

False positives were also observed in this study, where normal fetuses were misdiagnosed with cerebral-cranial anomalies. False-positive results lead to a detailed review by the sonographer, and if a definitive diagnosis cannot be made, referral to a specialist may be necessary. Although this process helps to improve diagnostic accuracy, frequent referrals may increase the workload of specialists, especially in resource-constrained areas, and may prolong the waiting time of patients and affect the efficiency of healthcare services.
Referral centers have demonstrated high specificity and sensitivity in the detection of CNS abnormalities, and these centers are able to accurately identify the majority of patients who actually have CNS abnormalities. Despite the better detection in referral centers, false-negative diagnoses are still more common in rural areas and local health facilities due to resource and professional constraints.
For pregnant women, referral usually involves more tests and counseling, increasing the patient's healthcare costs. For less well-off families, this can be a heavy burden. And false-positive results may themselves cause anxiety and distress in patients, which may be exacerbated by further referrals and tests. Prolonged psychological stress can have a detrimental effect on both the pregnant woman and the family.
These false-positive cases mainly occur in the presence of similar abnormal features in the ultrasound image, such as localized areas of hyperechoic or irregular structures. Such situations require continuous optimization of the model performance and gradual accumulation of data feedback in real-world applications to ensure its adaptation to different environments. Therefore, an effective feedback mechanism is needed to continuously improve model performance to reduce false-positive diagnoses before our DL system can be widely used in clinical practice.

Although the deep learning model in this study shows significant potential in the prediction of fetal cerebral-cranial anomalies, it still faces some limitations. Firstly, the small size of the dataset and the small number of samples for certain types of anomalies may lead to model bias in these categories. Second, differences in ultrasound equipment and different manipulation practices may affect image quality, which in turn may affect the stability and accuracy of the model.
Future studies should aim to increase the diversity of training samples through data augmentation, incorporate multimodal imaging techniques to provide more multidimensional information, and double validate model predictions by combining them with input from clinical experts. In addition, a continuous monitoring mechanism can be established to regularly assess model performance and make necessary adjustments and optimizations to ensure diagnostic accuracy and reliability.
With the continuous optimization and clinical validation of the technology, deep learning models are expected to become an important auxiliary tool for prenatal diagnosis worldwide, facilitating the rational allocation of healthcare resources and improving the health of mothers and babies.

\section{Materials and Methods}

\subsection*{Filtering datasets}

In order to maximize the accuracy of predicting fetal cerebral-cranial anomalies and to remove non-essential content that potentially affects the prediction, additional filtering was applied to the dataset. In the complete fetal ultrasound dataset, each ultrasound image includes information about each fetal mother and the source of the image in addition to the fetal ontology part, which would lead to data leakage during model training and not getting a good training model. At the same time, some of the images also include all kinds of textual data measured on fetal cerebral-cranial anomalies and other non-fetal body images, such as DV-S, DV-D, blood flow velocity waveform maps, and other color Doppler technology measured data, which will have a positive effect on the doctor's judgment of fetal anomalies, but to get these data need additional instruments to measure, which undoubtedly increase the cost of ultrasound examination. Our model focuses on fetal images and achieves excellent performance without these additional data, which definitely reduces the cost of ultrasound examination significantly. We have cropped the non-fetal body parts of these ultrasound images.

\subsection*{Baseline model}

We adopt the ResNet34 model from Deep Learning Systems as the main architecture. ResNet34 is a convolutional neural network based on Residual Network (ResNet), which solves the problem of gradient vanishing in deep network training by introducing residual blocks. The model contains multiple residual blocks, each consisting of two convolutional layers with inputs added directly to the outputs via jump connections. The specific structure of ResNet34 consists of an initial 7x7 convolutional layer, four stages of residual blocks (with feature maps numbering 64, 128, 256, and 512 respectively), as well as a globally averaged pooling layer and a fully connected layer. By using the ResNet34 model, we were able to efficiently extract complex features from fetal brain cranial anomaly images and achieve high accuracy in different types of anomaly classification tasks.

\subsection*{Construction of the DL system}

In this study, we used Leave-One-Out Cross-Validation (LOOCV) to evaluate the performance of the models and select the optimal model for each fold. Leave-One-Out Cross-Validation is a special cross-validation method in which only one sample at a time is set aside as the validation set and all the remaining samples are used as the training set. This method is particularly effective when the dataset is small, as it maximizes the use of available data and reduces the variance of the model.
In each leave-one-out  cross-validation, we trained a model and evaluated its performance on the set-aside validation set. We chose the model that performed best on the validation set as the optimal model for that fold. Eventually, we saved all these optimal models and combined them into an integrated model. We use Averaging in the integrated learning approach to derive the prediction results.
Through this integrated approach, we not only take advantage of the diversity of multiple models, but also improve the stability and generalization of the models. Averaging reduces the variance of individual models, making the final prediction results more reliable. In addition, this method is highly tractable and computationally efficient in practical applications for many types of machine learning.

\subsection*{Dividing dataset}
To ensure the accuracy and rigor of the prediction, we adopted a strict data partitioning strategy in leave-one-out cross-validation. Specifically, in the fetal cerebral-cranial anomalies dataset, the entire ultrasound image data of each pregnant woman is only used as a test set in any cross-validation and does not appear in the training set. This means that the training and test sets of each cross-validation are completely independent, and the data in the test set never participated in the training process, thus effectively avoiding the risk of data leakage.
In this way, we ensure the rigor and reliability of model performance. The test set for each cross-validation consists of all the ultrasound images of one pregnant woman, which do not appear in the training set. Ultimately, all folds yielded predictions of fetal cranial anomalies for each pregnant woman were averaged as the final result of the model. In the fetal cerebral-cranial normal dataset, we used a similar approach to divide the dataset. Specifically, for each of the four sectioning datasets, spinal longitudinal section, lateral ventricle transverse section, thalamus transverse section, and cerebellum transverse section, we carefully divided the samples into each sectioning dataset. Each type of slice dataset was divided into five pieces, each containing ultrasound images from a different pregnant woman, while ensuring that each pregnant woman only appeared in one piece and not in two or more pieces, thereby reducing potential bias and maintaining data independence.
This approach not only effectively ensures the performance of the model in practical applications, but also ensures that the model can accurately predict whether there is an abnormality in the fetal cranium of current pregnant women based on a large number of samples of fetal cranial abnormality ultrasound images from historical pregnant women.

Through rigorous cross-validation and data partitioning strategies, we improved the generalization ability and prediction accuracy of the model, making it more useful in clinical applications. In addition, this approach helps to reduce the risk of overfitting the model and improve its performance on new data. In the analysis of ultrasound images of fetal cerebral-cranial anomalies, data independence and rigor are particularly important. The ultrasound image data of each pregnant woman contains rich anatomical and pathological information, and by ensuring the independence of these data during the training and testing phases, we can more accurately assess the performance of the model. Our data segmentation strategy not only improves the predictive accuracy of the model, but also ensures its reliability and usefulness in real clinical applications. With this approach, we provide strong support for the early diagnosis and intervention of fetal cerebral-cranial anomalies.

\subsection*{Data pre-training processing}

During the data preprocessing stage, all fetal cranial brain ultrasound images were uniformly resized to 224 × 224 pixels to ensure consistent image sizes for input to the model. The training set images were randomly cropped, randomly flipped horizontally, converted to tensor and normalized to increase data diversity and model robustness. On the validation set, we wanted to preserve as much of the original characteristics of the images as possible in order to more accurately assess the generalization ability of the model. If the image is directly cropped to 224 × 224 pixels, some important edge information may be lost. Therefore, scaling the image to a slightly larger size (224 × 1.143 pixels) first, followed by center cropping, can better preserve the details and structure of the image. The mean and standard deviation used in the normalization process are [0.485, 0.456, 0.406] and [0.229, 0.224, 0.225] respectively, which are derived from the ImageNet dataset, and help to accelerate the convergence of the model and improve generalization.

\subsection*{Loss function}

Fetal cerebral-cranial anomaly ultrasound image datasets are inherently difficult to obtain, and as a result, there are some categories that are too few, resulting in unbalanced data. For example, there are too few anencephalic anomaly categories relative to other anomaly categories in our dataset. To deal with the category imbalance problem and to improve the classification performance of the model, we used a Weighted Cross-Entropy Loss function. Specifically, we first calculate the number of samples in each category and compute the category weights based on these numbers.
The class weights are calculated using the following formula:

$$
\textbf{class weights}_i = \frac{\textbf{total samples}}{\textbf{num classes} \times \textbf{class counts}_i}
$$

where:
\begin{itemize}
    \item $\textbf{total samples}$ is the total number of samples in the dataset,
    \item $\textbf{num classes}$ is the total number of classes,
    \item $\textbf{class counts}_i$ is the number of samples in class $i$.
\end{itemize}

This ensures that classes with fewer samples have higher weights, compensating for class imbalance. In this way, the categories with less number of samples are given higher weights and thus receive more attention during the training process. Eventually, we convert the computed category weights into a tensor and pass it as a parameter to the cross-entropy loss function for weighted loss computation. This weighted loss function helps to balance the influence of different categories and improve the performance of the model on a small number of categories.

\subsection*{Transfer learning}
Transfer learning is an effective method to improve prediction performance on new tasks by utilizing previously trained models \cite{li2020transfer}. In our experiments, we employ the ResNet34 model with its pre-trained weights on the large-scale ImageNet dataset. These pre-training weights contain a rich representation of image features and can significantly improve the model's performance on a variety of computer vision tasks. By using the pre-trained weights, we not only accelerate the convergence process of the model but also improve the generalization ability and robustness of the model, especially when the amount of data is limited. In addition, the pre-training weights help to reduce the risk of overfitting, which leads to better results in the task of classifying cranial anomalies in the fetal brain.

\subsection*{Training details}

In our study, the ‘best model’ performance is the set of optimal models for each fold based on the leave-one-out cross-validation. All models were trained using the AdamW optimizer with an initial learning rate of 0.0005 and a weight decay of 0.05. In order to improve the training stability and convergence speed of the models, we adopted a learning rate scheduling strategy and used a Warmup strategy in the early stages of training. We used a linear Warmup strategy within the first  epoch to gradually increase the learning rate from 0 to a preset initial learning rate of 0.0005. After the Warmup phase, the learning rate continued to be adjusted according to the cosine annealing strategy. This strategy helps the model to find a suitable optimization direction in the early stage of training, reduces the risk of gradient explosion, and accelerates convergence. 
Our experimental environment is configured as follows: Intel Xeon Gold 6138 CPU @ 2.00GHz, 96GB of RAM, and an NVIDIA RTX 2080 Ti GPU with 11GB of video memory. The system runs the Windows operating system and utilizes the PyTorch framework for development.
To ensure that the model achieves optimal performance on the validation set, we evaluate the validation accuracy of the model at the end of each epoch. If the validation accuracy of the current epoch is higher than the previously recorded optimal validation accuracy, the optimal validation accuracy is updated. If the validation accuracy does not improve for consecutive epochs, we trigger the early stop mechanism to stop the training process. This early stop mechanism helps prevent overfitting and saves computational resources.

%\section{List of Supplementary Materials}

%\vspace*{0.5 cm}

%\section*{Acknowledgment}
%Please insert text here.

%\begin{thebibliography}{1}
%\end{thebibliography}

\vspace*{ 1 cm}

\bibliographystyle{ieeetr}
\bibliography{references}

\begin{thebibliography}{10}

\bibitem{chitty1991effectiveness}
L.~S. Chitty, G.~H. Hunt, J.~Moore, and M.~O. Lobb, ``Effectiveness of routine ultrasonography in detecting fetal structural abnormalities in a low risk population.,'' {\em British Medical Journal}, vol.~303, no.~6811, pp.~1165--1169, 1991.

\bibitem{sonek2007first}
J.~Sonek, ``First trimester ultrasonography in screening and detection of fetal anomalies,'' in {\em American Journal of Medical Genetics Part C: Seminars in Medical Genetics}, vol.~145, pp.~45--61, Wiley Online Library, 2007.

\bibitem{xiao2023application}
S.~Xiao, J.~Zhang, Y.~Zhu, Z.~Zhang, H.~Cao, M.~Xie, and L.~Zhang, ``Application and progress of artificial intelligence in fetal ultrasound,'' {\em Journal of Clinical Medicine}, vol.~12, no.~9, p.~3298, 2023.

\bibitem{rajkomar2019machine}
A.~Rajkomar, J.~Dean, and I.~Kohane, ``Machine learning in medicine,'' {\em New England Journal of Medicine}, vol.~380, no.~14, pp.~1347--1358, 2019.

\bibitem{suzuki2017overview}
K.~Suzuki, ``Overview of deep learning in medical imaging,'' {\em Radiological physics and technology}, vol.~10, no.~3, pp.~257--273, 2017.

\bibitem{liu2019deep}
S.~Liu, Y.~Wang, X.~Yang, B.~Lei, L.~Liu, S.~X. Li, D.~Ni, and T.~Wang, ``Deep learning in medical ultrasound analysis: a review,'' {\em Engineering}, vol.~5, no.~2, pp.~261--275, 2019.

\bibitem{krizhevsky2012imagenet}
A.~Krizhevsky, I.~Sutskever, and G.~E. Hinton, ``Imagenet classification with deep convolutional neural networks,'' {\em Advances in neural information processing systems}, vol.~25, 2012.

\bibitem{yap2017automated}
M.~H. Yap, G.~Pons, J.~Marti, S.~Ganau, M.~Sentis, R.~Zwiggelaar, A.~K. Davison, and R.~Marti, ``Automated breast ultrasound lesions detection using convolutional neural networks,'' {\em IEEE journal of biomedical and health informatics}, vol.~22, no.~4, pp.~1218--1226, 2017.

\bibitem{yu2017deep}
Z.~Yu, E.-L. Tan, D.~Ni, J.~Qin, S.~Chen, S.~Li, B.~Lei, and T.~Wang, ``A deep convolutional neural network-based framework for automatic fetal facial standard plane recognition,'' {\em IEEE journal of biomedical and health informatics}, vol.~22, no.~3, pp.~874--885, 2017.

\bibitem{chen2017ultrasound}
H.~Chen, L.~Wu, Q.~Dou, J.~Qin, S.~Li, J.-Z. Cheng, D.~Ni, and P.-A. Heng, ``Ultrasound standard plane detection using a composite neural network framework,'' {\em IEEE transactions on cybernetics}, vol.~47, no.~6, pp.~1576--1586, 2017.

\bibitem{graves2013speech}
A.~Graves, A.-r. Mohamed, and G.~Hinton, ``Speech recognition with deep recurrent neural networks,'' in {\em 2013 IEEE international conference on acoustics, speech and signal processing}, pp.~6645--6649, Ieee, 2013.

\bibitem{gong2019fetal}
Y.~Gong, Y.~Zhang, H.~Zhu, J.~Lv, Q.~Cheng, H.~Zhang, Y.~He, and S.~Wang, ``Fetal congenital heart disease echocardiogram screening based on dgacnn: adversarial one-class classification combined with video transfer learning,'' {\em IEEE transactions on medical imaging}, vol.~39, no.~4, pp.~1206--1222, 2019.

\bibitem{arnaout2021ensemble}
R.~Arnaout, L.~Curran, Y.~Zhao, J.~C. Levine, E.~Chinn, and A.~J. Moon-Grady, ``An ensemble of neural networks provides expert-level prenatal detection of complex congenital heart disease,'' {\em Nature medicine}, vol.~27, no.~5, pp.~882--891, 2021.

\bibitem{connors2008fetal}
S.~L. Connors, P.~Levitt, S.~G. Matthews, T.~A. Slotkin, M.~V. Johnston, H.~C. Kinney, W.~G. Johnson, R.~M. Dailey, and A.~W. Zimmerman, ``Fetal mechanisms in neurodevelopmental disorders,'' {\em Pediatric neurology}, vol.~38, no.~3, pp.~163--176, 2008.

\bibitem{lin2022use}
M.~Lin, X.~He, H.~Guo, M.~He, L.~Zhang, J.~Xian, T.~Lei, Q.~Xu, J.~Zheng, J.~Feng, {\em et~al.}, ``Use of real-time artificial intelligence in detection of abnormal image patterns in standard sonographic reference planes in screening for fetal intracranial malformations,'' {\em Ultrasound in Obstetrics \& Gynecology}, vol.~59, no.~3, pp.~304--316, 2022.

\bibitem{xie2020using}
H.~Xie, N.~Wang, M.~He, L.~Zhang, H.~Cai, J.~Xian, M.~Lin, J.~Zheng, and Y.~Yang, ``Using deep-learning algorithms to classify fetal brain ultrasound images as normal or abnormal,'' {\em Ultrasound in Obstetrics \& Gynecology}, vol.~56, no.~4, pp.~579--587, 2020.

\bibitem{chowdhury2023stackfbas}
A.~A. Chowdhury, S.~H. Mahmud, K.~K.~S. Hoque, K.~Ahmed, F.~M. Bui, P.~Lio, M.~A. Moni, and F.~A. Al-Zahrani, ``Stackfbas: Detection of fetal brain abnormalities using cnn with stacking strategy from mri images,'' {\em Journal of King Saud University-Computer and Information Sciences}, vol.~35, no.~8, p.~101647, 2023.

\bibitem{paladini2007sonographic}
D.~Paladini, G.~Malinger, A.~Monteagudo, G.~Pilu, I.~Timor~Tritsch, A.~Toi, {\em et~al.}, ``Sonographic examination of the fetal central nervous system: guidelines for performing the ‘basic examination’and the ‘fetal neurosonogram’,'' {\em Ultrasound in Obstetrics \& Gynecology}, vol.~29, no.~1, pp.~109--116, 2007.

\bibitem{paladini2018ultrasound}
D.~Paladini and P.~Volpe, {\em Ultrasound of congenital fetal anomalies: differential diagnosis and prognostic indicators}.
\newblock CRC press, 2018.

\bibitem{nicolaides1986ultrasound}
K.~Nicolaides, S.~G. Gabbe, S.~Campbell, and R.~Guidetti, ``Ultrasound screening for spina bifida: cranial and cerebellar signs,'' {\em The Lancet}, vol.~328, no.~8498, pp.~72--74, 1986.

\bibitem{kunpalin2021cranial}
Y.~Kunpalin, J.~Richter, N.~Mufti, J.~Bosteels, S.~Ourselin, P.~De~Coppi, D.~Thompson, A.~David, and J.~Deprest, ``Cranial findings detected by second-trimester ultrasound in fetuses with myelomeningocele: a systematic review,'' {\em BJOG: An International Journal of Obstetrics \& Gynaecology}, vol.~128, no.~2, pp.~366--374, 2021.

\bibitem{li2020transfer}
X.~Li, Y.~Grandvalet, F.~Davoine, J.~Cheng, Y.~Cui, H.~Zhang, S.~Belongie, Y.-H. Tsai, and M.-H. Yang, ``Transfer learning in computer vision tasks: Remember where you come from,'' {\em Image and Vision Computing}, vol.~93, p.~103853, 2020.

\end{thebibliography}

\end{document}